# Magnetization Dynamics, Bennett Clocking and Associated Energy Dissipation in Multiferroic Logic

Mohammad Salehi Fashami[1], Kuntal Roy[2], Jayasimha Atulasimha[1*] and Supriyo Bandyopadhyay[2]

[1]Department of Mechanical and Nuclear Engineering, [2]Department of Electrical and Computer Engineering, Virginia Commonwealth University, Richmond, VA 23284, USA.

It has been recently shown that the magnetization of a multiferroic nanomagnet, consisting of a magnetostrictive layer elastically coupled to a piezoelectric layer, can be rotated by a large angle if a tiny voltage of few tens of mV is applied to the piezoelectric layer. The potential generates stress in the magnetostrictive layer and rotates its magnetization by $\sim 90^0$ to implement Bennett clocking in nanomagnetic logic chains. Because of the small voltage needed, this clocking method is far more energy-efficient than those that would employ spin transfer torque or magnetic fields to rotate the magnetization. In order to assess if such a clocking scheme can be also reasonably fast, we have studied the magnetization dynamics of a multiferroic logic chain with nearest neighbor dipole coupling using the Landau-Lifshitz-Gilbert (LLG) equation. We find that clock rates of $\sim$ 2 GHz are feasible while still maintaining the exceptionally high energy-efficiency. For this clock rate, the energy dissipated per clock cycle per bit flip is $\sim$52,000 $kT$ at room temperature in the clocking circuit for properly designed nanomagnets. Had we used spin transfer torque to clock at the same rate, the energy dissipated per clock cycle per bit flip would have been ~ $4\times10^8$ *kT*, while with current transistor technology we would have expended ~ $10^6$ *kT*. For slower clock rates of 1 GHz, stress-based clocking will dissipate only ~ 430 *kT* of energy per clock cycle per bit flip, while spin transfer torque would dissipate about $10^8$ *kT*. This shows that multiferroic nanomagnetic logic, clocked with voltage-generated stress, can emerge as a very attractive technique for computing and signal processing since it can be several orders of magnitude more energy-efficient than current technologies.

---
[*] Corresponding author. E-mail: jatulasimha@vcu.edu

**I. Introduction**

There is significant interest in implementing digital logic circuitry with single domain nanomagnets instead of traditional transistors since the latter are believed to be energy-inefficient. Transistors switch by moving electrical charge into or out of their active regions. If this process is carried out non-adiabatically, then it dissipates an amount of energy equal to at least *NkTln(1/p)*, where *N* is the number of electrons (information carriers) moved into or out of the device, *T* is the temperature and *p* is the "bit error probability" associated with random switching [1, 2]. On the other hand, if logic bits are encoded in two stable magnetization orientations along the easy axis of an anisotropic single-domain magnet (or the single domain magnetostrictive layer of a multiferroic nanomagnet), then switching between these orientations can take place by dissipating only ~ *kTln(1/p)* of energy, regardless of the number of spins (information carriers) in the nanomagnet [2]. This is a remarkable result and accrues from the fact that exchange interaction between spins makes all the ~ $10^4$ spins in a single-domain nanomagnet behave collectively like a giant *single* spin [2, 3] (a single information carrier) and rotate in unison [2]. As a result, for the same bit error probability *p*, the ratio of the minimum energy that must be dissipated to switch a nanomagnet to that dissipated to switch a nanotransistor will be $\sim 1/N \ll 1$. This makes the nanomagnet *intrinsically* more energy-efficient. It should be understood however that it is the *interaction* between spins, which is normally absent between charges – and not any inherent advantage of spin over charge, or magnetism over electricity – that gives the nanomagnet the advantage.

Because of this innate advantage, nanomagnet based computing architectures are attracting increasing attention. In one version of nanomagnetic logic (NML) known as "magnetic quantum cellular automata"[†], Boolean logic gates are configured by placing nanomagnets in specific geometric patterns on a surface so that dipole interactions between neighbors elicit desired logic operations on the bits encoded in the magnetization orientations of the nanomagnets [4, 5]. This is exactly the same approach that was envisioned earlier for the Single Spin Logic (SSL) paradigm, where exchange interaction between spins played the role of dipole interaction between magnets, while up- and down-spin polarizations encoded the two logic bits [6]. In fact, there is a one-to-one correspondence between NML and SSL if we view a single domain nanomagnet as a giant spin.

Unfortunately, NML paradigms share a debilitating drawback with SSL. Since there is no isolation between the input and output bits (unlike in transistors), unidirectional propagation of logic bits from one stage to the next requires sequential *clocking* of the nanomagnets (much like in bucket-brigade devices and charge coupled device shift registers) [7, 8]. This is accomplished with *Bennett clocking* [9] which is implemented by forcibly rotating a nanomagnet's magnetization through ~$90^0$ from the easy to the hard axis prior to a bit propagating through it. That places the nanomagnet temporarily at its (unstable) energy maximum, so that when the propagating logic bit reaches it, the dipole interaction of the neighbors nudges the magnet to the right energy minimum (correct orientation along the easy axis) and thus propagates the logic bit unidirectionally [10, 11].

Bennett clocking in NML can be implemented in two different ways: either with a global agent (e.g. a global magnetic field) which simultaneously resets the magnetization of every nanomagnet in a logic chain along the hard axis prior to propagating a bit [5], or with a local agent (e.g. a local spin polarized current that exerts a spin transfer torque on each nanomagnet [10, 11]) which rotates every magnet's magnetization individually to align along the hard axis. The disadvantage of the global approach proposed in [5] is that it makes the computing architecture *non-pipelined* and hence impractically slow as well as error-prone [12], while the disadvantage of the local approach is that it requires individual access (e.g. electrical access for injecting spin polarized current) to each nanomagnet. Since non-pipelined architectures are too sluggish for practical use, only the local clocking approach is considered.

---

[†] This is actually a misnomer. This architecture performs Boolean logic operations and does not really behave as cellular automata.

Recently, we showed that local clocking of NML can be implemented by applying a small voltage to a nanomagnet made of multiferroics [13]. Such a nanomagnet consists of two elastically coupled piezoelectric and magnetostrictive layers as shown in Figure 1. An applied voltage generates strain in the piezoelectric layer which is transferred almost entirely to the magnetostrictive layer by elastic coupling if the latter layer is much thinner than the former [14, 15]. This strain/stress can cause the magnetization of the magnetostrictive layer to rotate by a large angle [16], which has been demonstrated in recent experiments, although not at the nanoscale [17]. These rotations are sufficiently large to fulfill the requirements of Bennett clocking in logic chains [13]. In the specific configuration discussed in this paper, the voltage strains the piezoelectric layer via the $d_{31}$ coupling and we ensure that uniaxial tension or compression is always applied along the y-axis by mechanically restraining the PZT layer from expansion or contraction along the x-axis (in-plane direction orthogonal to y-axis). The same could have been achieved by applying the electric field along the y-direction direction, which will generate a stress along it via the $d_{33}$ coupling/

In ref. [13], we considered Bennett clocking of logic chains where the logic switches are ellipsoidal multiferroic nanomagnets of major axis = 105 nm and minor axis = 95 nm. Each multiferroic nanomagnet is composed of 10 nm thick Ni (magnetostrictive) layer and 40 nm thick lead zirconium titanate, or PZT, (piezoelectric) layer that are elastically coupled. We showed that by applying a tiny voltage (~200 mV) across the PZT layer of such a device, one can generate more than enough stress to rotate the magnetization of the Ni layer by nearly $90^0$ from the major (easy) axis and align close to the minor (in-plane hard) axis [13]. Upon releasing the stress, the magnetization of the Ni layer flips and the logic switch changes its bit state from 0 to 1, or vice versa. This implements Bennett clocking. We also showed [see the supplementary material accompanying ref. [13]] that replacing Ni with a material that has better magneto-mechanical coupling (e.g. Terfenol-D) will reduce the voltage required from ~ 200 mV to ~16 mV. This will allow Bennett clocking of NML with a voltage pulse of amplitude ~ 16 mV [13], resulting in extremely low energy dissipation. That is a rewarding outcome since Bennett clocking does no useful computation itself (it merely steers a logic bit from one stage to the next) and therefore should consume as little energy as possible.

The results in ref. [13] were based on a *time-independent* analysis predicated on energy arguments and did not yield the switching delay associated with bit flips and the resulting clock rate. In this paper, we have solved the Landau-Lifshitz-Gilbert (LLG) equation to find these switching delays. Fortunately, they are *not* impractically long, which bodes well for multiferroic logic and shows that this logic family can be extremely energy-efficient while at the same time being fast. This is the contribution of this paper.

The rest of this paper is organized as follows. In Section II, we formulate the LLG equations for simulating the three dimensional transient magnetization dynamics in a chain of four dipole-coupled multiferroic elements (Terfenol-D/PZT) forming an NML logic wire, when the two elements in the center of the chain are stressed abruptly with a voltage to rotate their magnetizations by nearly $90^0$ for Bennett clocking. This allows us to study the temporal evolution of the magnetization orientations of each multiferroic magnet in the chain. In Section III, we study the dynamics associated with different modes of applying stress (application of a small compressive stress (5.2 MPa) followed by removal, application of a large compressive stress (40 MPa) followed by reversal to an equally large tensile stress, etc.) We show that by using a sequence of compressive and tensile stresses, we can switch a Terfenol-D/PZT multiferroic nanomagnet in a logic chain in ~0.5 ns while dissipating ~ $2.17 \times 10^{-16}$ Joules (52,000 *kT*) of energy per clock cycle per bit flip. This will allow the speed of Bennett clock to reach 2 GHz. If we had to obtain the same clock speed while using spin transfer torque to rotate the magnetization of a nickel nanomagnet[‡], then the energy dissipation would have been ~ $4 \times 10^8 \, kT$ per clock cycle per bit flip [18]. Therefore, spin transfer torque is far more wasteful of energy than stress, and by inference, ordinary NML is far less energy-efficient than *multiferroic* NML. Finally, in Section IV, we present our conclusions and identify the directions for future research.

**II. Magnetization Dynamics in a Dipole Coupled Chain of Multiferroic Nanomagnets Stressed with Local Electrostatic Potentials**

Consider a linear chain of ellipsoidal multiferroic nanomagnets, each of which has an inhomogeneous magnetization $\vec{M}(\vec{r})$ [19]. Such a chain is shown in Fig. 2. The dimensions of each nanomagnet are ~ 101.75 nm × 98.25 nm × 10 nm. In that case, the exchange coupling penalty precludes the formation of multi-domain states [19] so that we can ignore the spatial variation of magnetization within each magnetostrictive layer, and model it as a single-domain nanomagnet [3]. Consequently, we will be concerned only with the variation in the magnetization of any multiferroic nanomagnet (viewed as a giant classical spin) with time, under the influence of an effective magnetic

---

[‡] Nickel is more favorable to spin transfer torque than Terfenol-D since it has a much lower resistivity and hence dissipates much less power due to the flow of spin polarized current through it.

field $\vec{H}_{eff}$ as described by the Landau-Lifshitz-Gilbert (LLG) equation [19, 20]:

$$\frac{d\vec{M}}{dt} = -\gamma \vec{M} \times \vec{H}_{eff} - \frac{\alpha\gamma}{M_s}\left[\vec{M} \times \left(\vec{M} \times \vec{H}_{eff}\right)\right] \quad (1)$$

where $\vec{H}_{eff}$ is the effective magnetic field on any one multiferroic element, which is the derivative of the total energy of that element with respect to its magnetization [19]. Accordingly,

$$\vec{H}_{eff} = -\frac{1}{\mu_0 \Omega} \frac{\partial E}{\partial \vec{M}} \quad (2)$$

where $\mu_0$ is the permeability of vacuum and $E$ is the total free energy (not energy density), of a particular multiferroic element of volume $\Omega$ in the chain shown in Figure 2. The total free energy of any element in this chain is given by:

$$E = E_{dipole} + E_{stress-anisotropy} + E_{shape-anisotropy} \quad (3)$$

where $E_{dipole}$ is the dipole-dipole interaction energy due to interaction between nearest neighbors, $E_{shape-anisotropy}$ is the shape anisotropy energy due to the elliptical shape of the multiferroic element, and $E_{stress-anisotropy}$ is the stress anisotropy energy caused by the stress transferred to the magnetostrictive layer of the multiferroic upon application of an electrostatic potential to the piezoelectric layer. We assume that the magnetostrictive layer is polycrystalline so that we can neglect magnetocrystalline anisotropy.

Let us focus on two adjacent multiferroic elements in the chain (labeled as the $i^{th}$ and $j^{th}$ element), whose magnetizations have polar and azimuthal angles of $\theta_i, \phi_i$ and $\theta_j, \phi_j$ respectively (see Fig. 1 for definition of polar and azimuthal angles). The dipole-dipole interaction energy is:

$$E_{dipole}^{i-j} = \frac{\mu_0 M_s^2 \Omega^2}{4\pi R^3} \sum_{\substack{i-1 \\ j \neq i}}^{i+1} \left[-2(\sin\theta_i \cos\phi_i)(\sin\theta_j \cos\phi_j) + (\sin\theta_i \sin\phi_i)(\sin\theta_j \sin\phi_j) + \cos\theta_i \cos\theta_j\right] \quad (4)$$

where $M_s$ is the saturation magnetization, $\Omega$ is the volume of each magnetostrictive layer and $R$ is the separation between their centers.

The shape anisotropy energy of the $i^{th}$ element $E_{shape-anisotropy}$ is given by:

$$E_{shape-anisotropy} = -\int_\Omega \left(\frac{\mu_0}{2} \vec{M} \bullet \vec{H}_M \right) dV \tag{5}$$

$$\vec{H}_M = -[N_{d\_xx} M_x \hat{i} + N_{d\_yy} M_y \hat{j} + N_{d\_zz} M_z \hat{k}] \tag{6}$$

$$\begin{aligned} E_{shape-anisotropy} &= \left(\frac{\mu_0}{2}\right)\Omega[N_{d\_xx} M_x^2 + N_{d\_yy} M_y^2 + N_{d\_zz} M_z^2] = \\ &\left(\frac{\mu_0}{2}\right)(M_s^2 \Omega)[N_{d\_xx}(\sin\theta_i \cos\phi_i)^2 + N_{d\_yy}(\sin\theta_i \sin\phi_i)^2 + N_{d\_zz}(\cos\theta_i)^2] \end{aligned} \tag{7}$$

where $N_{d\_xx}, N_{d\_yy}$ and $N_{d\_zz}$ are respectively the demagnetization factors along the x-, y- and z-directions.

Note that

$$N_{d\_xx} + N_{d\_yy} + N_{d\_zz} = 1 \tag{8}$$

We will view the magnetostrictive layer as an ellipsoid whose major and minor axes diameters are *a* and *b*, and the thickness is *t*. In that case, the demagnetization factors are [21]:

$$\begin{aligned} N_{d\_yy} &= \frac{\pi}{4}\left(\frac{t}{a}\right)\left[1 - \frac{1}{4}\left(\frac{a-b}{a}\right) - \frac{3}{16}\left(\frac{a-b}{a}\right)^2\right] \\ N_{d\_xx} &= \frac{\pi}{4}\left(\frac{t}{a}\right)\left[1 + \frac{5}{4}\left(\frac{a-b}{a}\right) + \frac{21}{16}\left(\frac{a-b}{a}\right)^2\right] \\ N_{d\_zz} &= 1 - \frac{\pi}{4}\left(\frac{t}{a}\right)\left[2 + \left(\frac{a-b}{a}\right) + \frac{18}{16}\left(\frac{a-b}{a}\right)^2\right] \end{aligned} \tag{9}$$

provided *a>b*, *a/b*~1 and *a, b >> t*.

The stress anisotropy energy in the $i^{th}$ element due to a stress applied along its major axis is:

$$E_{stress-anisotropy} = -\frac{3}{2}[\lambda_s \sigma \Omega]\sin^2\theta_i \sin^2\phi_i \tag{10}$$

where $(3/2)\lambda_s$ is the saturation magnetostriction and the stress $\sigma$ is considered negative for compression and positive for tension.

Using equations (3) – (10), we can express the total energy of the $i^{th}$ element interacting with its nearest neighbors ($j^{th}$ nanomagnet) as:

$$E_i = \underbrace{\frac{\mu_0 M_s^2 \Omega^2}{4\pi R^3} \sum_{\substack{i-1 \\ j \neq i}}^{i+1} \left[ -2(\sin\theta_i \cos\phi_i)(\sin\theta_j \cos\phi_j) + (\sin\theta_i \sin\phi_i)(\sin\theta_j \sin\phi_j) + \cos\theta_i \cos\theta_j \right]}_{E_{dipole}}$$
$$+ \underbrace{\left(\frac{\mu_0}{2}\right)(M_s^2 \Omega)\left(N_{d\_xx}(\sin\theta_i \cos\phi_i)^2 + N_{d\_yy}(\sin\theta_i \sin\phi_i)^2 + N_{d\_zz}(\cos\theta_i)^2\right)}_{E_{shape-anisotropy}} + \underbrace{\left(-\frac{3}{2}\lambda_s \sigma_i \Omega\right)\sin^2\theta_i \sin^2\phi_i}_{E_{stress-anisotropy}}$$

(11)

where we have removed all terms that do not have a dependence on magnetization orientation $\theta, \phi$ since they do not affect $\vec{H}_{eff}$. This total energy is used to find $\vec{H}_{eff}$ from Equation (2).

We can simplify Equation (1) by normalizing the magnetization with respect to $M_s$ (saturation magnetization) which is a conserved quantity (constant) for a single domain magnetostrictive layer at a constant temperature.

This yields

$$\vec{m} = \frac{\vec{M}}{M_s}; \quad m_x^2 + m_y^2 + m_z^2 = 1. \tag{12}$$

Here $m_x$, $m_y$ and $m_z$ are respectively the x-, y- and z-components of the normalized magnetization vector $\vec{m}$.

With this approximation, the vector LLG equation in Equation (1) simplifies to three coupled scalar equations:

$$\frac{dm_x(t)}{dt} = \gamma \left( H_{eff-z}(t)m_y(t) - H_{eff-y}(t)m_z(t) \right)$$
$$\quad - \alpha\gamma \left( H_{eff-y}(t)m_x(t)m_y(t) - H_{eff-x}(t)m_y^2(t) - H_{eff-x}(t)m_z^2(t) + H_{eff-z}(t)m_x(t)m_z(t) \right)$$

$$\frac{dm_y(t)}{dt} = \gamma \left( H_{eff-x}(t)m_z(t) - H_{eff-z}(t)m_x(t) \right) \tag{13}$$
$$\quad - \alpha\gamma \left( H_{eff-z}(t)m_y(t)m_z(t) - H_{eff-y}(t)m_z^2(t) - H_{eff-y}(t)m_x^2(t) + H_{eff-x}(t)m_x(t)m_y(t) \right)$$

$$\frac{dm_z(t)}{dt} = \gamma \left( H_{eff-y}(t)m_x(t) - H_{eff-x}(t)m_y(t) \right)$$
$$\quad - \alpha\gamma \left( H_{eff-x}(t)m_z(t)m_x(t) - H_{eff-z}(t)m_x^2(t) - H_{eff-z}(t)m_y^2(t) + H_{eff-y}(t)m_y(t)m_z(t) \right)$$

where $H_{eff-j}$ is the j-th component of $\vec{H}_{eff}$.

Note that $m_x(t)$, $m_y(t)$ and $m_z(t)$ are not independent of each other as they are related through

Equation (12) and we can use the parametric representation:

$$m_x(t) = \sin\theta(t)\cos\phi(t) \; ; \; m_y(t) = \sin\theta(t)\sin\phi(t) \; ; \; m_z(t) = \cos\theta(t) \tag{14}$$

This simplifies Equation (13) to two coupled equations for the magnetization orientation $\theta_i$, $\phi_i$ for the *i*-th nanomagnet:

$$\begin{aligned}
\frac{dm_x^i(t)}{dt} &= \cos\phi_i(t)\cos\theta_i(t)\frac{d\theta_i(t)}{dt} - \sin\theta_i(t)\sin\phi_i(t)\frac{d\phi_i(t)}{dt} = \gamma(H_{eff-z}^i(\sin\theta_i(t)\sin\phi_i(t)) - H_{eff-y}^i\cos\theta_i(t)) - \\
&\quad \alpha\gamma(H_{eff-y}^i(\sin\theta_i(t)\cos\phi_i(t))(\sin\theta_i(t)\sin\phi_i(t)) - H_{eff-x}^i(\sin\theta_i(t)\sin\phi_i(t))^2 - H_{eff-x}^i(\cos\theta_i(t))^2 + \\
&\quad H_{eff-z}^i(\sin\theta_i(t)\cos\phi_i(t))(\cos\theta_i(t)) \\
\frac{dm_y^i(t)}{dt} &= \sin\phi_i(t)\cos\theta_i(t)\frac{d\theta_i(t)}{dt} + \sin\theta_i(t)\cos\phi_i(t)\frac{d\phi_i(t)}{dt} = \gamma(H_{eff-x}^i\cos\theta_i(t) - H_{eff-z}^i\sin\theta_i(t)\cos\phi_i(t)) - \\
&\quad \alpha\gamma(H_{eff-z}^i(\sin\theta_i(t)\sin\phi_i(t))(\cos\theta_i(t)) - H_{eff-y}^i(\cos\theta_i(t))^2 - H_{eff-y}^i(\sin\theta_i(t)\cos\phi_i(t))^2 + \\
&\quad H_{eff-x}^i(\sin\theta_i(t)\cos\phi_i(t))(\sin\theta_i(t)\sin\phi_i(t))
\end{aligned} \tag{15}$$

The above result shows that there are two independent degrees of freedom $\theta_i$, $\phi_i$ for each nanomagnet and they are influenced by coupling to neighboring nanomagnets whose magnetization orientations $\theta_j, \phi_j$ influence the $H_{eff}$ terms through dipole coupling as can be seen from the next few equations.

Using Equation (2), the components of $\vec{H}_{eff}^i$ can be written as:

$$\begin{aligned}
H_{eff-x}^i(t) &= -\frac{1}{\mu_0 M_s \Omega}\frac{\partial E_i(t)}{\partial m_x(t)} = -\frac{1}{\mu_0 M_s \Omega}\frac{\partial E_i(t)}{\partial(\sin\theta_i(t)\cos\phi_i(t))} \\
H_{eff-y}^i(t) &= -\frac{1}{\mu_0 M_s \Omega}\frac{\partial E_i(t)}{\partial m_y(t)} = -\frac{1}{\mu_0 M_s \Omega}\frac{\partial E_i(t)}{\partial(\sin\theta_i(t)\sin\phi_i(t))} \\
H_{eff-z}^i(t) &= -\frac{1}{\mu_0 M_s \Omega}\frac{\partial E_i(t)}{\partial m_z(t)} = -\frac{1}{\mu_0 M_s \Omega}\frac{\partial E_i(t)}{\partial(\cos\theta_i(t))}
\end{aligned} \tag{16}$$

Using Equation (11) in Equation (16), we can write the components of the effective magnetic field for each element as:

$$H_{eff-x}^i(t) = \left(\frac{M_s\Omega}{4\pi R^3}\right)\sum_{\substack{i-1 \\ j\neq i}}^{i+1}[2\sin\theta_j(t)\cos\phi_j(t)] - M_s(N_{d-xx})\sin\theta_i(t)\cos\phi_i(t)$$

$$H_{eff-y}^i(t) = -\left(\frac{M_s\Omega}{4\pi R^3}\right)\sum_{\substack{i-1 \\ j\neq i}}^{i+1}[\sin\theta_j(t)\sin\phi_j(t)]$$

$$-M_s(N_{d-yy})\sin\theta_i(t)\sin\phi_i(t) + \left(\frac{3}{\mu_0 M_s}\lambda_s\right)\sigma_i(t)\sin\theta_i(t)\sin\phi_i(t)$$

$$H_{eff-z}^{i}(t)=-\left(\frac{M_{s}\Omega}{4\pi R^{3}}\right)\sum_{\substack{i-1 \\ j\neq i}}^{i+1}[\cos\theta_{j}(t)]-M_{s}(N_{d-zz})\cos\theta_{i}(t) \tag{17}$$

Substitution of Equation (17) in Equation (15) allows us to compute the temporal evolution of the magnetization vector of any multiferroic element (i.e. the temporal evolution of $\theta_i$ and $\phi_i$) in the chain of Fig 2.

### III. Results and Discussions

We have used 4$^{th}$ order *Runge-Kutta* method to solve the system of coupled differential equations in Equations (15) and (17) for the linear chain of four coupled multiferroic elements shown in Fig. 2. The solution yields the orientation $\theta_i(t)$, $\phi_i(t)$ of the magnetization vector in any multiferroic element in the chain at any instant of time *t*.

In this study, we have assumed that the magnetostrictive layers are made of polycrystalline Terenol-D which has the following parameters: $(3/2)\lambda s = 9\times10^{-4}$, $M_s = 0.8\times10^6$ A/m [21, 22], and average Young's modulus $Y = 8\times10^{10}$ Pa [23]. We assume that the Gilbert damping constant for Terfenol-D is $\alpha = 0.1$ based on high $[\alpha > 0.1]$ values for heavier elements such as dysprosium [24]. The dimensions of each nanomagnet are ~ 101.75 nm × 98.25 nm × 10 nm and the center-to-center separation between neighboring elements (or pitch) is 200 nm. The above parameters were chosen to ensure that: (i) the shape anisotropy energy of the elements is sufficiently high (~0.8 eV or ~32$kT$ at room temperature) so that the equilibrium bit error probability due to spontaneous magnetization flipping is very low (~$e^{-32} \approx 10^{-14}$), (ii) the dipole interaction energy is limited to 0.26 eV which is significantly lower than the shape anisotropy energy to prevent spontaneous flipping of magnetization, but is still large enough to ensure that the magnetization of the multiferroic elements always flips to the correct orientation when stress is released, even under the influence of random thermal fluctuations. We recognize however that quantifying the relationship between switching speed, temperature, dipole coupling and error probability is beyond the scope of this work since that would need solving the stochastic LLG equation [25] or Fokker-Plank equations [19] rather than the deterministic equation in Equation (1).

In all our simulations, the initial magnetizations of the multiferroics always corresponds to the ground state of the array where the four magnetizations are anti-ferromagnetically ordered, i.e. each multiferroic's magnetization is along the major axis (which is the easy axis) and nearest neighbors have anti-parallel magnetizations as shown in the first row of Fig. 2. At time $t = 0$, the first multiferroic (far left) has its magnetization flipped abruptly (second row of Fig. 2). We then consider the time evolutions of the magnetizations of every multiferroic nanomagnet in various cases when stress is applied to the second and third nanomagnets in arbitrary time sequences.

The maximum value of stress that we have considered is 40 MPa which can be generated by a voltage of ~ 200 mV applied across the PZT layer. We calculate this as follows: The PZT layer can transfer up to $500 \times 10^{-6}$ strain to the Terfenol-D layer. This strain generates a stress of 40 MPa in the Terfenol-D layer, which is found by multiplying the strain with the average Young's modulus of Terfenol-D, assuming linearity. Since the piezoelectric coefficient of PZT $d_{31} \approx -10^{-10}$ m/V, the voltage required to induce this strain in the PZT layer that is 40 nm thick is 200 mV. The corresponding maximum stress-anisotropy energy is $\frac{3}{2}[\lambda_s \sigma \Omega] = 682$ $kT$, which is much more than the shape anisotropy energy barrier of 32 $kT$, and is therefore more than adequate to turn the magnetization to the hard axis from the easy axis. The excess energy of 650 $kT$ (682 $kT$ - 32$kT$) is consumed to speed up the rotation.

The local effective field on each nanomagnet $\vec{H}_{eff}(t)$ is calculated at each time step from Equation (17). We also assume that stress is applied instantaneously and removed instantaneously. The rationale for this assumption is that the capacitance of a 40 nm-thick PZT layer of surface area 101.75 nm × 98.25 nm is 1.74 fF, if we assume the relative dielectric constant of PZT to be 1000. We also assume that the PZT layer is electrically accessed with a silver wire of resistivity ~2.6 μΩ-cm [26] so that an access line of length 10 μm and cross section 50 nm × 50 nm has resistance ~100 Ω. Therefore, the RC time constant associated with charging the capacitor is 0.174 picoseconds while the magnetization switching time is always more than 0.5 nanoseconds. This allows us to consider the onset and removal of stress as instantaneous. Furthermore, the mechanical resonance frequency of such as system can be approximately calculated as $f = \frac{1}{4L}\sqrt{Y/\rho}$, where $\rho$ is the density and $L$ is the long dimension. Since, the PZT layer is much thicker than the Terfenol-D layer, we assume average $\rho$ = 7,500 Kg/m$^3$ [27], average Young's modulus dominated by PZT is $Y$=60 GPa [27] and $L$~100 nm. Consequently, the resonance frequency turns out to be 7 GHz. We may be able to scale the size to $L$ ~50 nm to increase the resonance frequency to ~ 14 GHz (that corresponds to a time period of 70 ps), which is shorter than 0.5 ns. Hence, it is a very good approximation to consider the stress to be applied instantaneously. This analysis shows that ultimately the Bennett clock rate of multiferroic logic is likely to be limited by the mechanical response of the structure!

We discuss two illustrative cases, with the first case being the simplest, in which the logic chain in Fig. 2 is Bennett clocked by applying only compressive stress of 5.2 MPa to the second and third nanomagnets. The voltage required to generate this stress is 26 mV (the voltage scales linearly with stress; hence, if 200 mV generates 40 MPa, then 26 mV will generate 5.2 MPa). The second and third nanomagnets are stressed instantaneously at times $t = 0$ and $t = 0.02$ ns respectively, assuming that the first nanomagnet's magnetization has been flipped by some external agent at $t = 0$ to provide input data to the chain. Once stress has rotated the second and third nanomagnets' magnetizations by nearly $90^0$ (i.e. their projections on the plane of the magnets have undergone a $90^0$ rotation to align along the common hard axis), it is removed abruptly from the second nanomagnet while still being held constant on the third. The relaxed second nanomagnet then gradually settles down to the correct magnetization state anti-parallel to that of the first because of the influence of its shape anisotropy and dipole interaction from its neighbors. This is shown in the fourth row of Fig. 2. The input bit, provided to the first nanomagnet, has now successfully propagated to the second, which means that Bennett clocking has been successfully implemented.

The simulation result in Fig. 3 shows that complete switching of the second multiferroic's magnetization vector (from "down" to "up") takes ~ 1 ns. Note that the switching corresponds to the azimuthal angle $\phi_2$ of the second magnet changing from $-90^0$ to $+70^0$. After the second nanomagnet has switched, we can release the stress on the third. Therefore, the stress on the third magnet needs to be maintained for a total duration of ~ 1 ns, which means that the maximum clock rate achievable in this case is $1/(1\text{ ns}) = 1$ GHz.

In the second case, we apply a larger 40 MPa compressive stress on the second and third multiferroics until their magnetizations align along the hard axis (i.e. $\phi_2$ becomes $0^0$). We then reverse the stress on the second multiferroic from compressive to tensile, which aids it to relax faster from the hard axis to the easy axis. As a result, the total switching time to switch the second multiferroic's magnetization vector reduces to ~ 0.5 ns as can be seen in Fig. 6. However, in this case, the high stress causes significant "out of plane" excursion of the magnetization vector. We discuss the two cases below.

Case 1: Compressive stress of 5.2 MPa is applied instantaneously on multiferroic nanomagnets 2 and 3 by applying a potential of 26 mV, followed by instantaneous removal of stress from multiferroic nanomagnet 2 after its magnetization aligns close to the in-plane hard axis.

We apply a compressive stress of 5.2 MPa on the second and third multiferroic nanomagnets as a step function in time at t=0 and t=0.02 ns respectively. Since, Terfenol-D has positive magnetostriction, this tends to rotate their magnetizations to a direction perpendicular to the direction of the applied stress. It should be noted that we assume that both magnetization orientations rotate to the right to simplify the numerical analysis. The analysis would be identical if both magnetizations rotated to the left, because of the symmetry. By "phasing" our clock so that stress is applied on the second nanomagnet slightly before it is applied on the third, we ensure that the x-component of $\vec{H}_{dipole}$ due to the initial rotation of the second nanomagnet favors lining up the third nanomagnet's magnetization in the same direction (parallel). Ultimately, the second magnet's magnetization turns anti clockwise from $\phi_2 = -90^o$ to nearly $0^o$ and third multiferroic's magnetization rotates clockwise from $\phi_3 = +90^o$ to nearly $0^o$, so that they both align close to the hard axis and are mutually parallel. As shown in Fig 3, the time taken for this $90^o$ rotation to occur, which orients the second and third multiferroics along the hard axis, is ~0.4 ns. At this point, the nearest-neighbor dipole coupling makes the first and fourth multiferroics' magnetizations rotate slightly away from the "down" orientation to the "down and slightly right" orientation so that their orientations are $\phi_1 \approx -82^0$, $\phi_4 \approx -66^0$. This is shown in Fig. 3. These peripheral multiferroic elements rotate because of dipole coupling even though no stress is applied on them. However, the dipole coupling is not strong enough to make them overcome their shape anisotropy energy, so they do not switch or flip their magnetizations.

After ~0.4 ns have elapsed and both the second and third multiferroics have their magnetizations oriented close to the in-plane hard axis, stress is removed abruptly from the second multiferroic, while the third is still held at 5.2 MPa compression. As shown in Fig.3, the magnetization of the second multiferroic now gradually relaxes to the nearly "up" state due to dipole interactions with its two neighbors and shape anisotropy. This shows successful execution of Bennett clocking, but this last relaxation takes another ~0.6 ns. Thus, the switching process that flips the second multiferroic's magnetization from "down" to nearly "up" takes a total time of ~ 0.4 ns + 0.6 ns = ~1.0 ns Hence, a bit propagates through one unit of the logic chain in ~ 1 ns, which makes the maximum allowed clock rate 1 GHz.

Let us now focus on the peripheral elements in the chain (multiferroics 1 and 4). After t~ 0.4 ns, the first element's magnetization begins to rotate back towards $\phi_1 = -90^0$ once stress in removed from the second element. However, it can never quite reach $\phi_1 = -90^0$ because the second element's magnetization does not rotate beyond $\phi_2 \approx 70^0$ owing to the strong x-component of $\vec{H}_{dipole}$ caused by the magnetization of the third element. This dipole field also causes the fourth element's magnetization to settle at $\phi_4 \approx -66^0$. Since we had ensured that the dipole energy is much smaller than the shape anisotropy energy, the peripheral elements cannot rotate beyond $\sim -65^0$.

As already stated, the voltage required to generate a stress of 5.2 MPa in the PZT layer is ~ 26 mV. Hence, the energy dissipated in the clocking cycle is $(1/2)CV^2$ = 140 $kT$ at room temperature during the turn-on phase of the voltage and another $(1/2)CV^2$ = 140 $kT$ during the turn-off phase.

Thus, by dissipating 280 *kT* of energy *in the clocking circuit*, we can achieve ~ 1 GHz clock rate.

There is however some additional energy dissipated in the magnet itself when it reverses magnetization [10]. This energy is calculated as [28]

$$E_d = \int_0^\tau \frac{\alpha \mu_0 \gamma \Omega}{(1+\alpha^2) M_s} \left| \vec{M} \times \vec{H}_{eff} \right|^2 dt \qquad (18)$$

where $\vec{M} \times \vec{H}_{eff}$ is the effective torque acting on a nanomagnet due to the combined effects of shape anisotropy, stress and dipole interaction [28]. This energy is calculated numerically for all four magnets following the prescription of ref. [28] and then added up. It turns out to be another ~ 150 *kT*. Thus, the total energy dissipated per clock cycle per bit flip in this case is ~ 430 *kT*.

The magnetization vector of any magnet of course need not be constrained to the plane of the magnet under stress. It can lift out of the plane and the out-of-plane excursion is measured by the polar angle $\theta$. Fig. 4 shows the extent of the out-of-plane excursion of the magnetization vector. The polar angles $\theta_2, \theta_3$ deviate by no more than $3^0$ from $90^0$, which is the magnet's plane, when the stress is 5.2 MPa. Thus, as long as the stress is small, the magnetization vector barely lifts out of the magnet's plane and virtually all the rotation takes place in the plane. The complex motion of the tip of the magnetization vector in three-dimensional space is shown in Fig. 5. Note that the tip always resides pretty much in the *x-y* plane which is the plane of the magnet. However, even the small out-of-plane excursion has a significant effect on the switching delay. It speeds up the switching because the "out-of-plane" magnetization leads to a significant $H_{eff}$ along the z-direction due to the large out-of-plane shape anisotropy (demagnetization factor $N_{d\_zz}$). Interestingly, this out of plane $H_{eff}$ provides a large torque $\left( \vec{M} \times \vec{H}_{eff} \right)$ that speeds up the in-plane rotation because switching via the precessional mode is faster than switching via the damped mode.

Case 2: Compressive stress of 40 MPa is applied on multiferroic nanomagnets 2 and 3 to align their magnetizations along the hard axis, followed by applying a tensile stress of 40 MPa on multiferroic nanomagnet 2 to help it relax to its easy axis faster and in the process flipping its magnetization.

The lessons learned from Case I tell us that we can make the switching process faster if we:

(a) increase the magnitude of stress on the nanomagnets since that will result in a larger "effective field" $\vec{H}_{eff}$, and

(b) Make the relaxation from the hard to the easy axis faster for the second nanomagnet. This relaxation is slow since the only "driving force" on the nanomagnet after stress is removed comes from the effective field produced by the shape anisotropy and dipole coupling. Consequently, application of a tensile stress that drives the magnetization away from the hard axis could increase the "driving force" and make the relaxation faster. This would require that we reverse the stress from compressive to tensile on the second nanomagnet (by reversing the polarity of the voltage) after its magnetization vector reaches the hard axis.

Fig. 6 shows that merely increasing the compressive stress on the second and third nanomagnets from 5.2 MPa to 40 MPa decreases the time it takes to align both nanomagnets along the hard axis to about ~ 0.1 ns from the ~0.5 ns found in Case I. Once nanomagnets 2 and 3 line up along their common hard axis, we reverse the sign of the stress on the second nanomagnet from 40 MPa compression to 40 MPa tension, which then makes the magnetization relax to the nearly "up" state in only another ~0.5 ns, after all the ripples and ringing die down to around ~5º from the easy axis. Thus, by increasing the magnitude of stress and by aiding the relaxation process with stress reversal, we can shorten the total switching time from 1 ns to about 0.5 ns. This increases the maximum clock rate from 1 GHz to 2 GHz.

Here again, the dipole coupling is not strong enough to overcome the shape anisotropy energy; therefore, the magnetizations of first and the fourth nanomagnets do not rotate beyond $\sim -65^0$. The magnetization of the first nanomagnet reaches $\sim -83^0$ around 0.15 ns but then rotates back to $-90^0$ (as the second nanomagnet settles close to the $+90^0$ state due to application of a high tensile stress). The x-component of the $\vec{H}_{dipole}$ due to the magnetization of the third nanomagnet makes the magnetization of the fourth rotate further and settle at $\sim -66^0$.

The energy dissipated in the clocking circuit is computed as follows: When the compressive stress is turned on, we will dissipate an energy of $(1/2)CV^2$ in the clock line attached to either nanomagnet 2 or nanomagnet 3. When stress is reversed, we will dissipate an additional energy of $(1/2)C(2V)^2$ in the clock line attached to magnet 2. Finally, when stress is removed, we will dissipate an energy of $(1/2)CV^2$ in the lines attached to either magnet. Thus, the total energy that we will spend to flip the magnetization of the second magnet is $3CV^2$, which is 50,000 *kT* since V = 200 mV and C = 1.74 fF.

To this energy we must add the energy dissipated in all four magnets during magnetization reversal. This additional energy is calculated numerically following the method of ref. [28] and it turns out to be another ~ 2000 kT at room temperature. Hence the total energy dissipated per clock cycle per bit flip is ~ 52,000 *kT*.

Fig. 7 shows the out-of-plane excursion of the magnetization vector. In this case, $\theta_2, \theta_3$ deviate by $\pm 15^0$ from the 90º position, showing that the magnetization vector lifts out of the magnet's plane by $\pm 15^0$. This produces a large out-of-plane $H_{eff}$ as explained earlier, which produces a large torque $\left(\vec{M} \times \vec{H}_{eff}\right)$ that speeds up the switching by causing significant precessional motion of the magnetization vector. Fig. 8 shows the complex dynamics of the tip of the magnetization vector in three-dimensional space. This complex dynamics is responsible for all the ripples we see in Fig. 6.

**IV. Conclusions**

In summary, we have studied the magnetization dynamics associated with Bennett clocking of multiferroic logic by formulating and solving the appropriate LLG equations. Our results show that clock rates of 2 GHz are achievable with proper design if we use common materials like Terfenol-D and lead zirconium titanate (PZT) to construct the multiferroic logic switches. For a clock rate of 2 GHz, the energy dissipated per clock cycle per bit flip can be 1-2 orders of magnitude smaller than in transistor circuits [31] and at least 3 orders of magnitude smaller than in NML clocked with spin transfer torque. On the other hand, if we are willing to settle for a clock rate of 1 GHz, then the energy dissipated is potentially 4 orders of magnitude smaller than in transistor circuits [31] and 6 orders of magnitude smaller than in NML driven with spin transfer torque.. Moreover, transistors tend to have a leakage current and hence encounter significant standby power dissipation, which NML does not. Therefore, NML employing multiferroic nanomagnets can emerge as a very viable candidate for the next generation of computers and signal processors.

**Appendix: A note on multiferroic materials**

While there are a few naturally occurring single-phase magnetoelectric (ME) materials, the magnetoelectric coupling they exhibit are either weak or occur at too low temperatures [15]. Most research has therefore focused on studying combinations of different piezoelectric and magnetostrictive material composites that are elastically coupled to produce the magnetoelectric effect. Initial work started with the development of magnetoelectric composites in 1970s by directional solidification of eutectic Fe-Co-Ti-Ba-O [29]. Later research showed that very high coupling is achieved by using Terfenol-D/PZT laminates [30]. This is because one would try to use piezoelectrics with large $d_{31}$ coefficients (such as PZT) to maximize the stress/strain transferred while choosing a material with large magnetostriction (such as Terfenol-D) to increase the stress-anisotropy energy generated for a given stress. Polycrystalline Ni [22] has saturation magnetostriction ~ $-30 \times 10^{-6}$. Nanoscale/thin-film polycrystalline FeGa [17] and Terfenol-D [23] would have saturation magnetostriction ~ $150 \times 10^{-6}$, ~ $900 \times 10^{-6}$ respectively, motivating the use of Terfenol-D for optimized device performance. Further, by extending this switching paradigm to multiferroic magnets with biaxial anisotropy [32], one can configure four state logic with applications beyond combinational and sequential circuits [33].

**Acknowledgement:** We thank Prof. Avik Ghosh and his graduate student Ms. Kamaram Munira at University of Virginia for discussions on $H_{eff}$ and for independently verifying our results.

**References**

[1] Cavin R K, Zhirnov V V, Hutchby J A and Bourianoff G I 2003 *Proc. IEEE.* **91** 1934.

[2] Salahuddin S and Datta 2007 *Appl. Phys. Lett.* **90** 093503.

[3] Cowburn R P, Koltsov D K, Adeyeye A O, Welland M E and Tricker D M 1999 *Phys. Rev. Lett.* **83** 1042.

[4] Cowburn R P and Welland M E 2000 *Science* **287** 1466.

[5] Csaba G, Imre A, Bernstein G H, Porod W and Metlushko V 2002 *IEEE Trans. Nanotech.* **1** 209.

[6] Bandyopadhyay S, Das B and Miller A E 1994 *Nanotechnology* **5** 113.

[7] Bandyopadhyay S and Roychowdhury V P 1996 *Japan J. Appl. Phys. Pt. 1* **35** 3350.

[8] Bandyopadhyay S 2005 *Superlatt. Miscrostruct.* **37** 77.

[9] Bennett C H 1982 *Int. J. Theor. Phys.* **21**, 905.

[10] Behin-Aein B, Salahuddin S and Datta S 2009 *IEEE Trans. Nanotech.* **8** 505.

[11] Behin-Aein B, Datta D, Salahuddin S and Datta S 2010 *Nature Nanotechnol.* **5** 266.

[12] Bandyopadhyay S and Cahay M 2009 *Nanotechnology* **20** 412001.

[13] Atulasimha J and Bandyopadhyay S 2010 *Appl. Phys. Lett.* **97** 173105.

[14] Eerenstein W, Mathur N D and Scott J F 2006 *Nature* **442** 17;
    Zheng H, et al., **303** (5658): 661-663, Science 2004.

[15] Nan C, Bichurin M I, Dong S, Viehland D and Srinivasan G 2008 *J. Appl. Phys.* **103**, 031101.

[16] Atulasimha J, Flatau A B and Cullen J R 2008 *J. of Appl. Phys.* **103**, 014901.

[17] Brintlinger T, Lim S-H, Baloch K H, Alexander P, Qi Y, Barry J, Melngailis J, Salamanca-Riba L, Takeuchi I and Cummings J 2010 *Nano Lett.* **10**, 1219.

[18] Ralph D C and Stiles M D 2008 J Magn. Magn. Mater. 2008 **320**, 1190.

[19] Bertotti G, Serpico C, and Mayergoyz I D 2008 *Nonlinear Magnetization Dynamics in Nanosystems (Elsevier Series in Electromagnetism),* Elsevier, Oxford

[20] Gilbert T L 2004 *IEEE Trans. Magn.* **40**, 3443.

[21] Chikazumi S 1964 *Physics of Magnetism* John Wiley & Sons, New York.

[22] Ried K, Schnell M, Schatz F, Hirscher M, Ludescher B, Sigle W and Kronmuller H 1998 *Phys. Stat. Sol. (a)* **167**, 195; Abbundi R and Clark A E 1977 *IEEE Trans. Mag.* **13** 1547.

[23] Kellogg R A and Flatau A B 2008 *J. of Intelligent Material Systems and Structures* **19** 583.

[24] Walowski J, Djorjevic Kaufman M, Lenk B, Hamann C, McCord J and Münzenberg M 2008 *J. Phys. D: Appl. Phys.* **41** 164016.

[25] D. E. Nikonov and G. I. Bourianoff and G. Rowlands and I. N. Krivorotov, 2010, J. Appl. Phys., **107**, 113910; Spedalieri F M, Jacob A P, Nikonov D and Roychowdhury V P 2009 arXiv:0906:5172v1 [cond-mat.mes-hall].

[26] http://www.itrs.net/Links/2009ITRS/2009Chapters_2009Tables/2009_Interconnect.pdf.

[27] http://www.piezo.com/prodmaterialprop.html.

[28] Roy K, Bandyopadhyay S and Atulasimha J 2011 arXiv:1101:2222v1 [cond-mat.mes-hall].

[29] Run A M J G, Terrell D R and Scholing J H 1974 *J. Mater. Sci* **9** 1710.

[30] Ryu J, Priya S, Carazo A V, Uchino K and Kim H-E 2001 *J. Amer. Ceramic Soc.* **84** 2905.

[31] CORE9GPLL_HCMOS9_TEC-4.0 Databook, ST Micro (2003).

[32] N. A. Pertsev and H. Kohlstedt, Appl. Phys. Lett., 95, 163503, (2009).

[33] N. D'Souza, J. Atulasimha, S. Bandyopadhyay, 2011, arXiv:1101.0980v1 [cond-mat.mes-hall]

**Figure Captions**

**Fig. 1:** A bi-layer multiferroic nanomagnet composed of a magnetostrictive layer and a piezoelectric layer.

**Fig. 2:** Propagating a logic bit through a chain of four dipole coupled multiferroic nanomagnets with Bennett clocking implemented with stress. (First row) a chain of elliptical nanomagnets in the ground state with magnetization orientation indicated by arrows. (Second row) Magnetization of the first magnet is flipped with an external agent and the second magnet finds itself in a tied state where it experiences no net dipole interaction. (Third row) The second and the third magnet are subjected to electrically induced stresses that rotate their magnetizations close to the hard axis. (Fourth row) The second magnet is freed from stress so that its magnetization relaxes to the easy axis as a result of shape anisotropy, and it switches to the desired "up" state rather than the incorrect "down" state since the dipole interaction from the left neighbor is now stronger than that from the right neighbor so that the tie is resolved. (NOTE: The coordinate system in the right corner should also be interpreted as showing the projection of the magnetization vector on the plane of the nanomagnet.)

**Fig. 3:** (a) Magnetization angles $\phi$ [which are the projections of the magnetization vector on the magnet's plane] versus time plotted for the four multiferroic nanomagnets (PZT/Terfenol-D) in the chain shown in Fig. 2 when compressive stresses of 5.2 MPa are applied abruptly to the second and third nanomagnets at time $t = 0$ and t=0.02 ns, respectively. . Stress is removed abruptly from the second nanomagnet after 0.386 ns when it assumes an orientation along the in-plane hard axis while the third nanomagnet remains stressed throughout this time interval. Note that even though magnets 1 and 4 are unstressed, their magnetizations rotate slightly because of dipole interaction with their stressed neighbors.

**Fig. 4:** Out-of-plane excursion of the magnetization vector. Polar angles $\theta$ versus time plotted for the four nanomagnets in the chain shown in Fig. 2 when the second and third nanomagnets are subjected to the stress cycle of Fig. 3.

**Fig. 5:** Three dimensional plot of magnetization components of nanomagnet 2 showing the spatial excursion of the tip of the magnetization vector. The stress cycle on all magnets is the same as in Figs. 3 and 4.

**Fig. 6:** Magnetization angle $\phi$ versus time plotted for the four nanomagnets in the chain of Fig. 2. A compressive stress of 40 MPa is applied abruptly on the second and third nanomagnets at time $t = 0$ and t=0.02 ns respectively with a voltage of 0.2 V. Stress on the second nanomagnet is reversed from compression to tension by switching the polarity of the voltage after 0.095 ns (i.e. after the nanomagnets come close to the hard axis) while the third nanomagnet is held at 40 MPa compression.

**Fig. 7:** Out-of-plane excursion of the magnetization vector. Polar angle $\theta$ versus time plotted for the four nanomagnets in the chain of Fig. 2.  The stress cycle is the same as in Fig. 6.

**Fig. 8:**  Three dimensional plot of magnetization components of nanomagnet 2 showing the spatial excursion of the tip of the magnetization vector. The stress cycle on all magnets is the same as in Figs. 6 and 7.

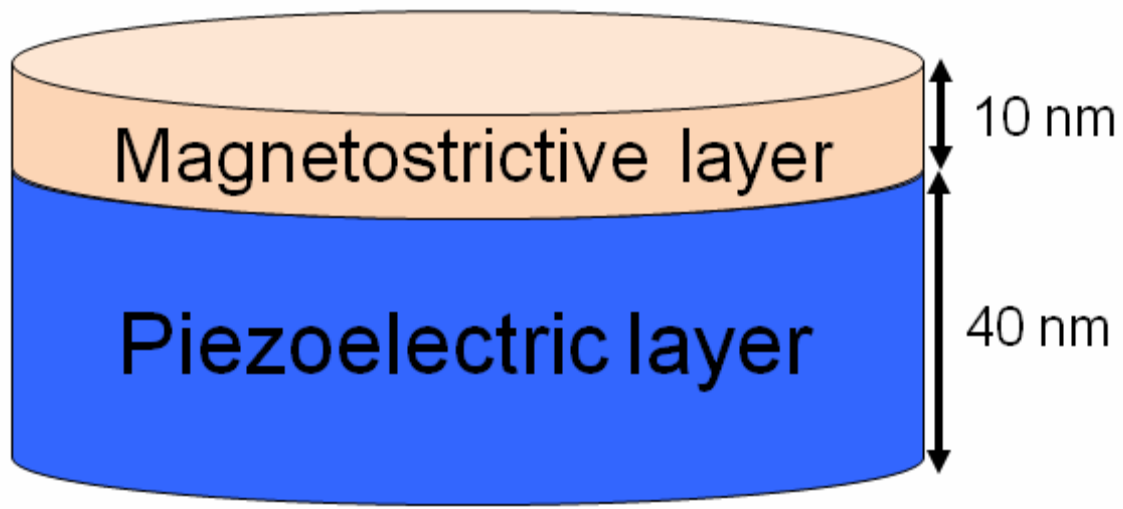

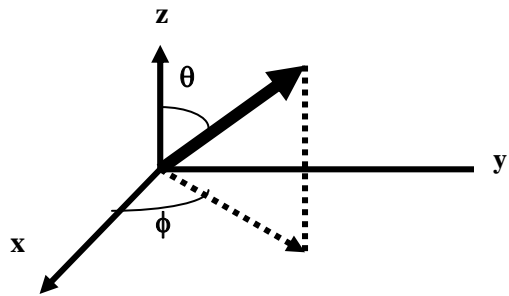

Fig. 1

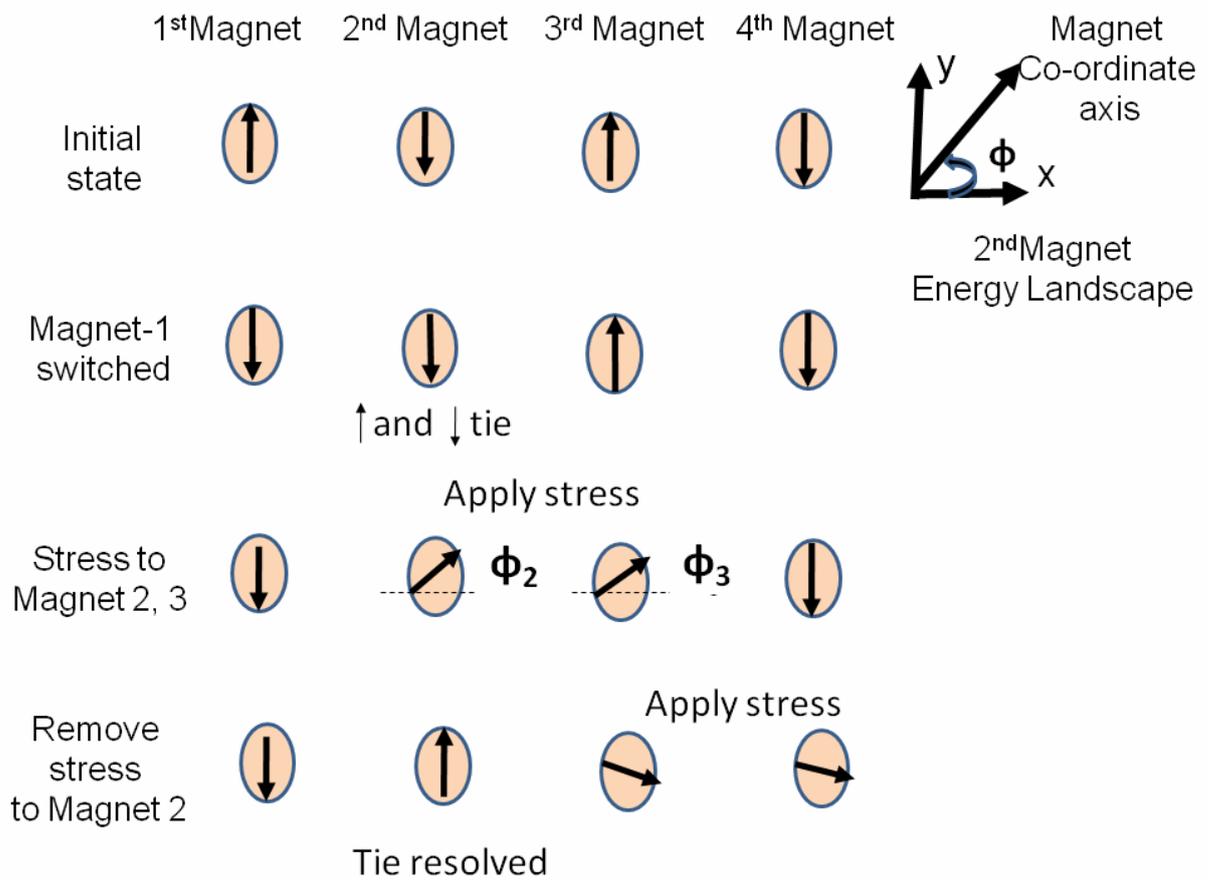

Fig. 2

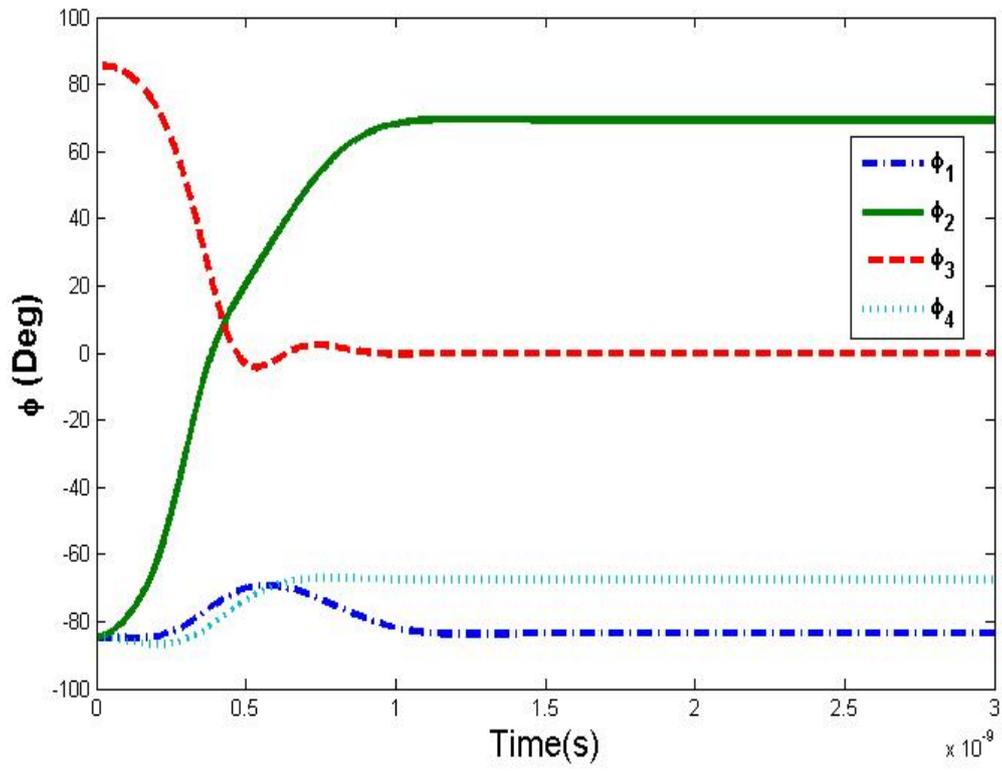

Fig. 3

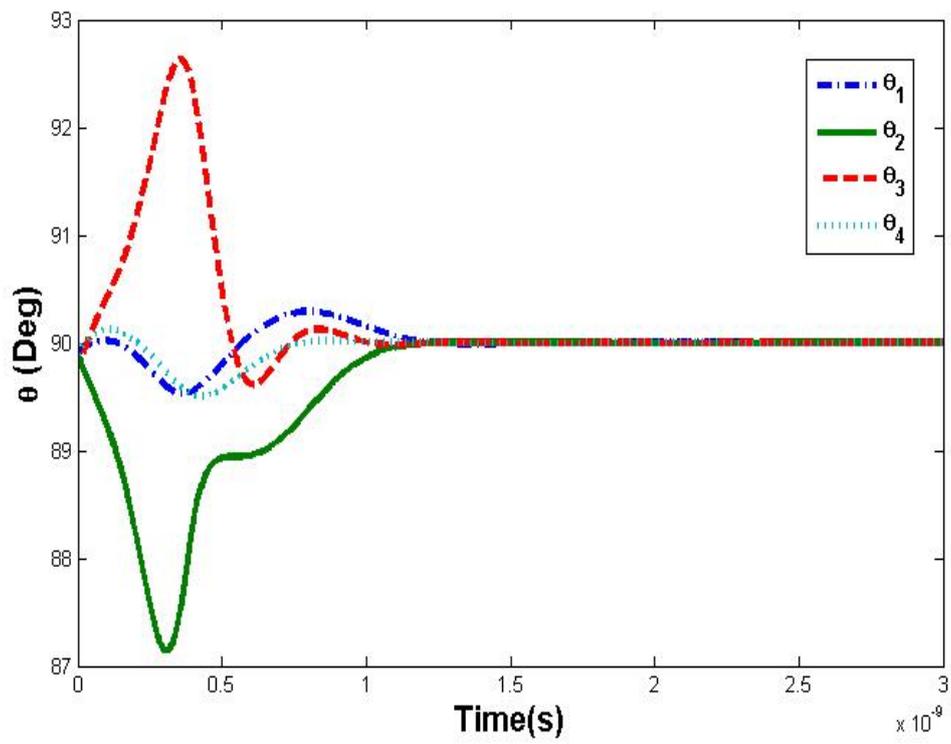

Fig. 4

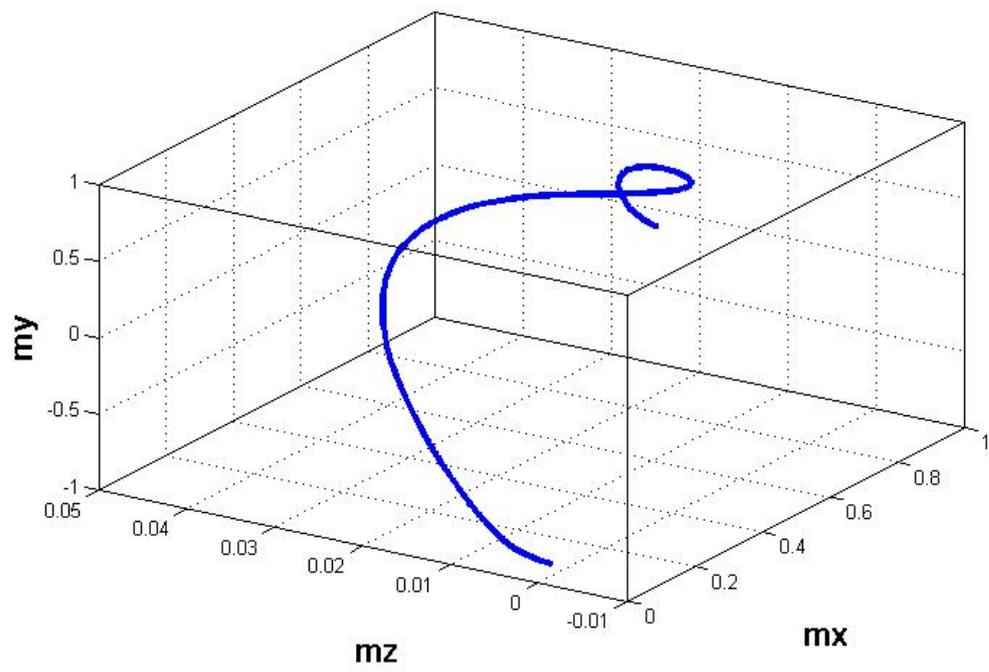

Fig. 5

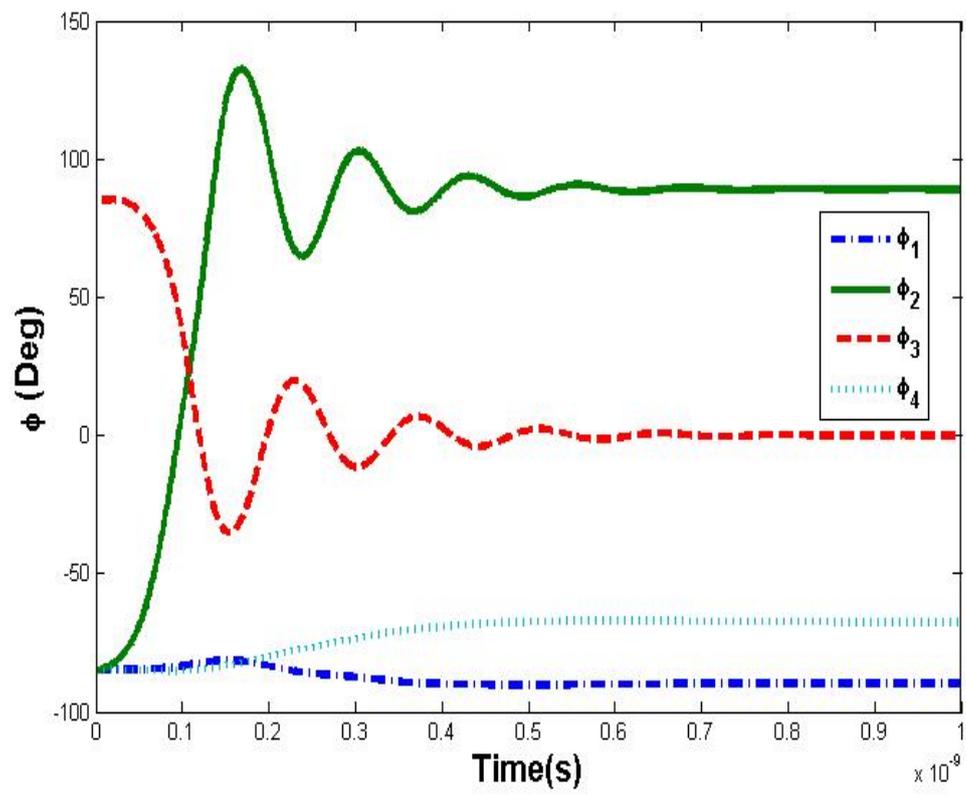

Fig.6

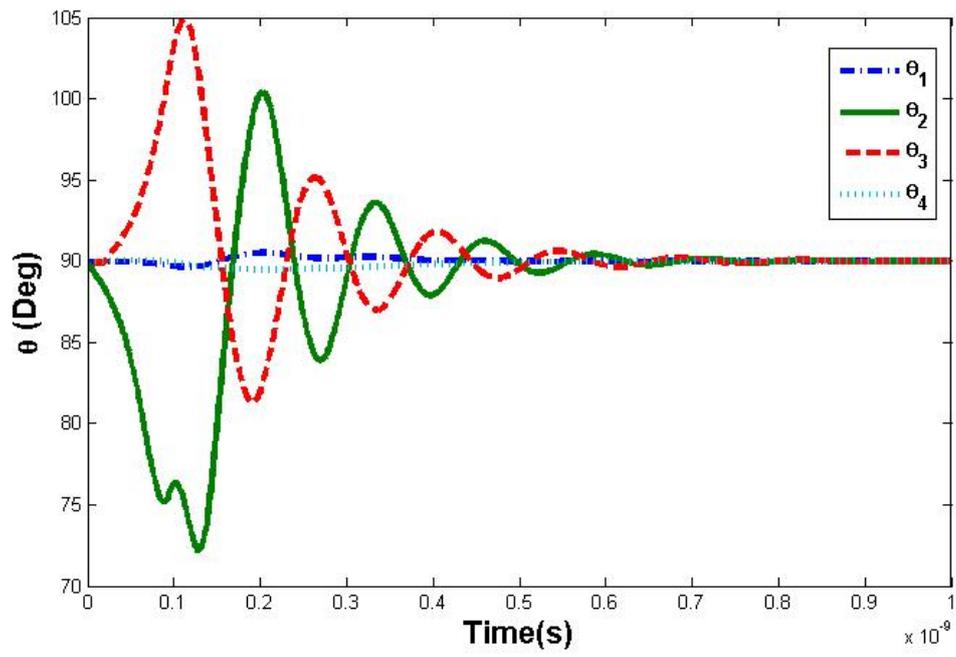

Fig. 7

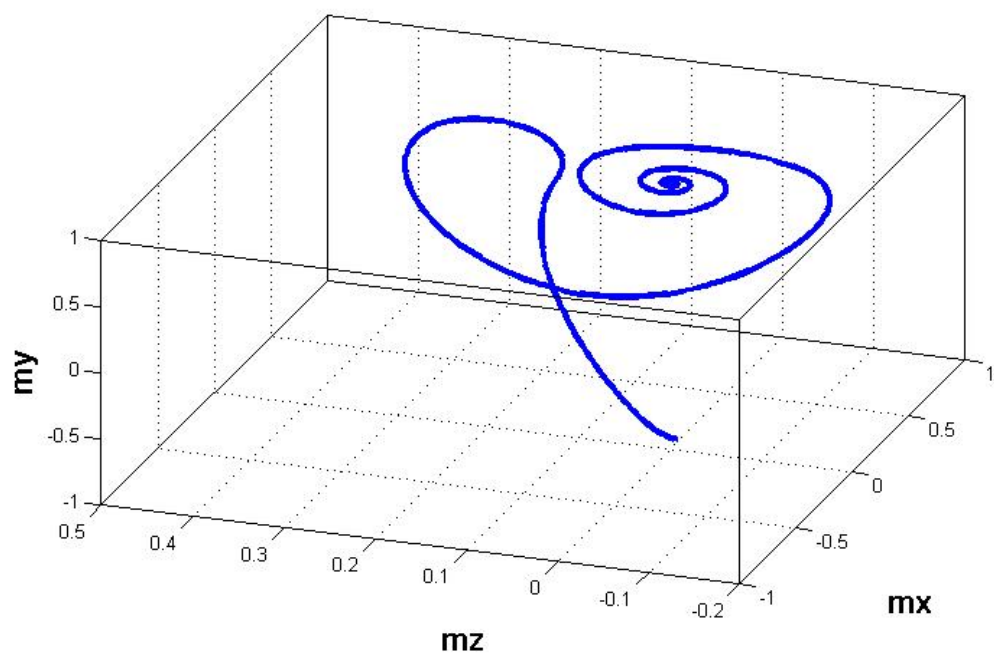

Fig. 8